\documentclass[a4paper,aps,pre,floatfix,nofootinbib,superscriptaddress,showpacs]{revtex4-1}

\usepackage[pdftex]{graphicx}
\usepackage{amsmath}    

\begin{document}

\title{
Predicting partially observed processes on temporal networks
by Dynamics-Aware Node Embeddings (DyANE)
}
\author{Koya Sato}
\affiliation{University of Tsukuba, Tokyo, Japan}
\author{Mizuki Oka}
\affiliation{University of Tsukuba, Tokyo, Japan}

\author{Alain Barrat}
\affiliation{Aix Marseille Univ, Universit\'e de Toulon, CNRS, CPT,  Marseille, France}
\affiliation{Tokyo Institute of Technology,  Tokyo,  {Japan}}
\author{Cattuto}
 \affiliation{University of Turin, {Turin}, {Italy}}
\affiliation{ISI Foundation, {Turin}, {Italy}}


\begin{abstract} 
Low-dimensional vector representations of network nodes have proven successful to feed graph data to machine learning algorithms and to improve performance across diverse tasks. Most of the embedding techniques, however, have been developed with the goal of achieving dense, low-dimensional encoding of network structure and patterns. Here, we present a node embedding technique aimed at providing low-dimensional feature vectors that are informative of dynamical processes occurring over temporal networks -- rather than of the network structure itself -- with the goal of enabling prediction tasks related to the evolution and outcome of these processes. We achieve this by using a modified supra-adjacency representation of temporal networks and building on standard embedding techniques for static graphs based on random-walks. We show that the resulting embedding vectors are useful for prediction tasks related to paradigmatic dynamical processes, namely epidemic spreading over empirical temporal networks. In particular, we illustrate the performance of our approach for the prediction of nodes' epidemic states in a single instance of a spreading process. We show how framing this task as a supervised multi-label classification task on the embedding vectors allows us to estimate the temporal evolution of the entire system from a partial sampling of nodes at random times, with potential impact for nowcasting infectious disease dynamics.
\end{abstract}

\maketitle

\section{Introduction}
\label{sec:intro}

The ubiquity of network representations of 
widely different systems has led to a flourishing
of methods aimed at the analysis of their structure
\cite{albert2002statistical,newman2003structure}
and of processes taking place on networks, 
such as information diffusion, epidemic spread, synchronization, etc
\cite{Barrat:2008,pastor2015epidemic}. 
Recently, these investigations have been extended to the case of temporal networks, in which nodes and links can appear and disappear in time \cite{Holme:2012,Holme:2015}.

Most works aim in particular at understanding how 
the network's features impact the outcome of processes taking place on top of them, usually considering
averages over many realizations of a stochastic process.
A less considered issue concerns the reconstruction of
partially observed processes taking place on a network. Indeed,
given a dynamical process occurring on a network, such that nodes change state over time, in a realistic setting
only partial knowledge of this evolution can in general be  envisioned, as for instance in diffusion processes such as the spread of contagious diseases or rumors. 
Recovering the complete information on the unfolding of the process from partial observations can then be of crucial importance, for instance
to estimate the actual impact of a spread whose evolution is only partially known, and whose parameters are a priori unknown. This issue has been addressed in the specific case of 
spreading processes, under various hypothesis.
For instance, some works have put forward  methods to recover the state of all nodes and the seeds of a spread from a partial observation of nodes at a given time~\cite{sundareisan2015hidden,xiao2018SteinerTreeSampling}, without attempting to recover the whole temporal evolution of the process. 
Methods to recover the state evolution of all nodes have also been proposed, using as input snapshots of the whole system, i.e., the knowledge of the state of all the nodes at a certain time~\cite{Sefer2016,8408489,8667709}. 
Finally, several methods using partially observed snapshots have also been proposed~\cite{Altarelli:2014b,rozenshtein2016reconstructing,xiao2018reconstructing,xiao2018SteinerTreeSampling}, 
typically based on strong assumptions on the nature of the underlying diffusion process. 

Here, we propose a novel approach to tackle the general issue of 
recovering all the information about a single instance of
a partially observed and unknown process, leveraging the recent development of node embedding methods. Network node embedding methods have indeed recently gained a lot of popularity~\cite{8294302,goyal2018graph,GOYAL2019} as tools
able to explore network structure, and we propose here that  embeddings can also be designed in order to recover infomation on dynamical processes on networks. 
In short, a node embedding maps each node of a network into a low-dimensional vector, such that the vectors representing different nodes are close if the network nodes share some
similarity or are close in the network. Node 
embeddings thus aim at exposing in the low-dimensional space
structural features and relevant patterns of the network that are not 
necessarily evident in the network representation.
Most importantly, the embedding vectors can be used as feature vectors in machine learning applications, and have been shown to yield improved performance for tasks such as node classification,
link prediction, clustering, or visualization. 
 
Here we show that node embedding methods can also be tailored to the study of \textit{dynamical processes on temporal networks},
and in particular to the task described above of predicting the evolution and outcome of one instance of the dynamics (e.g., an epidemic spread) from partial 
information and without detailed knowledge of the dynamical process itself.
A useful embedding should thus yield low-dimensional vectors that encode information relevant to the \textit{dynamics} of the process occurring over a temporal network -- rather than information about the network \textit{structure} itself.
Since dynamical processes unfold over time-respecting paths determined by the underlying network and by its evolution over time, we argue that the sought embeddings should be informative of these paths -- the paths along which information can propagate.
Driven by this idea, we propose to first map the temporal network to a static graph representation, a so-called supra-adjacency representation, whose nodes are the \texttt{(node,time)} pairs of the original temporal network~\cite{Valdano:2015}.
We modify the original supra-adjacency representation method to only consider nodes at those times when they interact,
and we map the original temporal edges to edges between the corresponding \texttt{(node,time)} pairs: this static graph representation preserves
the temporal paths of the original temporal network (i.e., the paths supporting and constraining the dynamical process at hand). An example of the supra-adjacency representation we use here is shown in Fig.~\ref{fig:flattening}.
Since the resulting representation is a static graph, we can then apply standard embedding techniques:
we focus on embeddings based on random walks 
\cite{Perozzi2014,Grover2016} 
as they provide an efficient way to sample the relevant paths.

We show the performance of the proposed embeddings for the prediction of partially observed processes 
in the context of a paradigmatic dynamical process -- epidemic spread over temporal networks -- in which network nodes exist in few discrete states and the dynamics consists of transitions between such states (e.g., a ``susceptible'' node becoming ``infectious''). As described above,
we focus on the task of predicting the nodes' states over time for a single realization of the epidemic process. 
Specifically, we set up a multi-label supervised classification problem with a training set obtained by sampling the node states at random times, with no information about the mechanics of state transitions nor on the parameters of the epidemic process. In summary, our contributions are as follows:
\begin{itemize}
    \item We propose a new method for node embedding tailored to the study of dynamical process on temporal networks, using a modified supra-adjacency representation for temporal networks and building on standard random-walk based embeddings for static graphs.
    \item We show that in the important case of epidemic spreading, a good prediction performance of nodes' states can be achieved in a supervised multi-label classification setting informed by the proposed embeddings.
    \item We show that our method achieves good performance in estimating the temporal evolution of the entire system from sparse observations, consistently across several data sets and across a broad range of parameters of the epidemic model. Our approach requires no fine-tuning of the embedding hyper-parameters and yields consistently superior performance than other embedding methods.
\end{itemize}

The paper is organized as follows: we first formulate in detail the problem at hand in Section \ref{sec:problem}. We then 
describe our approach in Section \ref{sec:approach}
and show the results of numerical experiments in Section 
\ref{sec:experiments}. We conclude with some 
perspectives in Section \ref{sec:conclusion}.

\section{Problem Formulation}
\label{sec:problem}

Let us first state in general terms 
the problem we want to address: 
given a known temporal network, an unknown process
unfolding on this network and a partial observation
of the dynamical states of the nodes, we want to 
predict the dynamical state of all nodes at all times. 
Crucially, this prediction must 
be performed without any information on the details
of the dynamical process taking place on the network, except
for the set of possible states of each node. In particular,
we do not make any assumption on the type of transitions,
the parameter values, nor even 
on the reversibility or irreversibility of the
process. Moreover, this prediction does not concern an average
over various realizations of the same process, but instead
one single realization, which is partially observed.

\subsection{Temporal network}

In more precise terms, we consider a temporal network $g$ in discrete time on 
a time interval $T = (1, 2, \cdots, |T|)$, 
i.e., a set $V$ of $N=|V|$ nodes 
and a set of temporal edges of the form $(i,j,t)$ denoting
that nodes $i$ and $j$ are in interaction at time $t \in T$.
Note that each temporal edge can also potentially carry a weight $w_{ij}(t)$. The set of temporal edges at $t$ is denoted $E_t$,
and $V_t$ is the set of nodes which have at least one temporal edge at $t$: 
the snapshot network at $t$ is the undirected
weighted network  $G_t=(V_t, E_t)$.

For each node $i$, we define its set of active times 
$T_{i}$ as the set of timestamps $t$ in which it is
involved in at least one temporal edge
(i.e., such that $i \in V_t$). 
We denote the $a$-th active time of $i$ by $t_{i, a} \in T_{i}$,
with $t_{i, a} < t_{i, a+1}$, and
we  define the set of active copies of each node $i$, 
that we call "active nodes",  as
$\mathcal{V}_{i} = \{(i, t)| t \in T_{i}\}$.
An active node is thus of the form \texttt{(node,time)}.
The overall set of active nodes is the union of all the sets
of active nodes,
i.e.,  $\mathcal{V} = \cup_{i \in V}\mathcal{V}_{i}$.

\subsection{Dynamical process}
We consider a dynamical process taking place on the weighted temporal network,
such that each node $i \in V$ can be at each time 
in one of a finite set of discrete states ${\mathcal S}$.
Nodes can change state either  spontaneously or through interaction along temporal edges.
Our definition is thus very general and encompasses in particular models of epidemic propagation, rumor propagation, opinion formation or cascading processes \cite{Barrat:2008,Castellano:2009,Pastor-Satorras:2015}.

While the problem description is very general and encompasses
a wide variety of processes on networks, we will focus here on a paradigmatic
dynamical process of strong relevance, namely the Susceptible-Infectious-Recovered (SIR) model for epidemic spreading, which is widely used to model contagious infections such as flu-like diseases~\cite{KeelingRohani}.
In this model, each node can be 
at each time in one of three possible states:
susceptible (S), infectious (I), and recovered (R). 
At the start of the process, all nodes are in state
S, except for the epidemic seeds, which are in 
state I. A contact between an S and an I nodes leads to a contagion event in which the S node becomes infectious
with probability $1-(1-\beta)^w$ at each timestamp, where $\beta$ is the infection rate and $w$ is the edge weight between the S and I nodes.
Let us denote by $I_t$ the set of infectious nodes at $t$, and
consider a susceptible node $i$. We denote its set of neighbours at $t$ as
 $N_{t}(i) = \{j| (i, j, t) \in E_t \}$, and
$N_{t}(i)\cap I_{t}$ is the set of its infectious neighbours at $t$. 
The probability that none of these infectious neighbours transmits the disease to $i$
at timestep $t$ is
$\prod_{j \in N_{t}(i)\cap I_{t}}{ (1-\beta)^{w(i, j, t)} } $
and thus the probability that $i$ becomes infectious 
at time $t$, due to its interactions, is 
$1-\prod_{j \in N_{t}(i)\cap I_{t}}{ (1-\beta)^{w(i, j, t)} }$.
Recovery from state I to state R occurs also stochastically: 
each infectious node becomes
recovered at each timestamp with probability
$\mu$. Recovered nodes do not change state any more. The parameters of the model are thus
the infection and recovery rates $\beta$ and $\mu$~\cite{KeelingRohani}.

We note here that the SIR model -- in addition to its relevance to many real-world phenomena -- is particularly interesting to study in the context of the prediction problem addressed
in this paper:
it features indeed not only state transitions occurring upon interaction (hence, along the edges of the temporal network) 
but also spontaneous state transitions 
that can occur at any time, and in particular 
between successive active times of a node (the infectious-recovered transition).

\subsection{Partial observation of the process and prediction task}

We assume that a sample of the dynamical evolution of the
process is known. More precisely, we first assume that 
the state of a node can only be observed when it is active, i.e., in contact with at least another node. 
Denoting by $f$: $(i, t) \in \mathcal{V} \to s \in {\mathcal S}$
the mapping that specifies the state of each node at each of its active times, we assume that this mapping is only 
partially known, through the observation of a fraction of the active nodes: we define the set of the 
observed active nodes as $D \subset \mathcal{V}$.
Here, for simplicity, we will assume that $D$ results from a uniform random sampling of $\mathcal{V}$.

The task at hand is then to predict the state of all the 
unobserved active nodes, i.e., the state in which each node
is at each of its active times. This allows to reconstruct
the unfolding of the process both at the local
node level and obviously as well at the population level.
In the example of the SIR process, crucial outcomes
of the prediction task are the epidemic curve, including
the timing of the epidemic peak, and the final epidemic
size, i.e., the actual number of nodes that have been 
affected by the spreading process.

\section{Our Approach: DyANE}
\label{sec:approach}
Our approach consists of three steps.
First, we map the temporal network 
to a static network between active nodes through
a modified supra-adjacency representation.
Second, we apply standard embedding techniques 
for static graphs to this supra-adjacency network. 
We will consider embeddings based on random walks as
they explore the temporal paths on which transmission between nodes can occur. 
Finally, we train a classifier to predict the dynamical state of all active nodes
based on the vector representation of active nodes and the partially observed states. We now give details on each of these
steps.

\begin{figure}[thb] 
    \includegraphics[width=\linewidth]{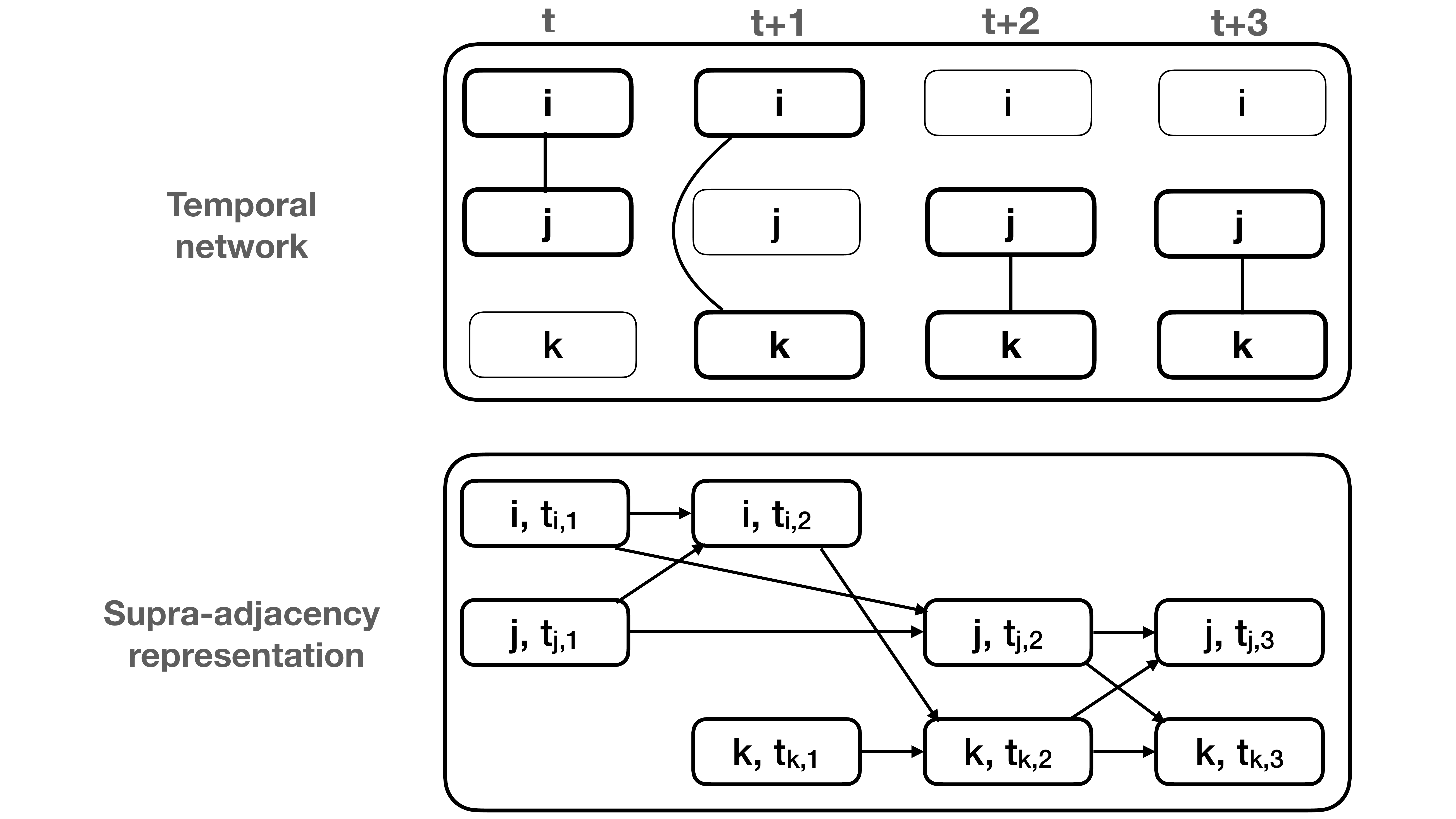} 
    \caption{ {Modified supra-adjacency representation (\textit{dyn-supra})}. 
    The top panel shows the original temporal network at four successive times. At each time $t$ we show in bold the 
    nodes of $V_t$, i.e., the nodes with at least one temporal edge. This network is  
      mapped to a static representation (bottom) where nodes are \texttt{(node,time)} pairs of the original network, keeping 
      for each node of the original network only the times in which it is active.}
    \label{fig:flattening} 
\end{figure}

\subsection{Supra-adjacency representation}
We first map the temporal network to 
a supra-adjacency representation. 
The supra-adjacency representation has been first
developed for multilayer networks~\cite{Gomez:2013,Kivela:2014}, in which nodes interact on different layers (for instance
different communication channels in a social network). 
It has been generalized to temporal networks,
seen as special multilayer networks in which every
timestamp is a layer \cite{Valdano:2015}:
each node of the supra-adjacency representation is identified by the pair of indices $(i,t)$, corresponding to the node label 
$i$ and the time frame $t$ of the original temporal network.
In this representation, the nodes $(i,t)$ are  
present for all nodes
$i$ and timestamps $t$, even if $i$ is isolated at $t$.

We propose here to use a modified version in which 
we consider only the active times of each node. This results
in a supra-adjacency representation whose nodes are the
active nodes of the temporal network.
More precisely, we define the supra-adjacency network as 
$\mathcal{G} = (\mathcal{V}, \mathcal{E})$, where 
$\mathcal{E}$ are (weighted, directed) edges joining active nodes.
The mapping from the temporal network to the supra-adjacency network 
consists of the following two procedures (Fig.~\ref{fig:flattening}):
\begin{itemize}
    \item For each node $i$, we connect its successive active versions:
    for each active time $t_{i, a}$ of $i$,
    we draw a directed ``self-coupling'' edge from
    $(i, t_{i, a})$ to $(i, t_{i, a+1})$
    (recall that active times are ordered in increasing temporal order). 
    
    \item For each temporal edge 
 $(i, j, t)$, the time $t$ corresponds by definition to an active time
 for both $i$ and $j$, that we denote respectively by
 $t_{i, a}$ and $t_{j, b}$.
 We then map 
 $(i, j, t) \in E$
 to two directed edges $\in \mathcal{E}$, namely 
 $\left((i, t_{i, a}), (j, t_{j, b+1})\right)$ 
 and 
 $\left((j, t_{j, b}), (i, t_{i, a+1})\right)$.
 In other words, the active copy of $i$ at $t$, $(i,t)$,
 is linked to the next active copy of $j$, and vice-versa.
\end{itemize}

The first procedure makes each active node adjacent to its nearest past and future versions (i.e., 
at the previous and next active times). This
ensures that a node carrying an information at a certain time
can propagate it to its future self along the self-coupling edges, and is useful in an embedding perspective to favor temporal continuity.
The second procedure encodes the temporal interactions,
and yields the crucial property that any time-respecting path existing on the original temporal network, 
on which a dynamical process can occur, 
is also represented in the supra-adjacency representation.
Indeed, if an interaction between two nodes
$i$ and $j$ occurs at time $t$ and 
potentially modifies their states, e.g.,
by contagion or opinion exchange or modification, 
this can be observed and will have consequences
only at their next respective active times:
for instance, if $i$ transmits a disease to $j$ 
at $t$, $j$ can propagate it further to other
neighbours only at its next active time, 
and not immediately at $t$.
This is reflected in the supra-adjacency representation we propose.

The edges in $\mathcal{E}$ are thus of two types,
joining two active nodes corresponding either to
the same original node, or to distinct ones.
For each type, we can consider various ways of assigning
weights to the edge.
We first consider for simplicity that all self-coupling edges
carry the same weight $\omega$, which becomes thus a
parameter of the procedure. 
Moreover, we simply report the 
weight $w_{ij}(t)$ of each original temporal edge 
$(i,j,t)$ on the two supra-adjacency edges
$\left((i, t_{i, a}), (j, t_{j, b+1})\right)$ 
 and 
 $\left((j, t_{j, b}), (i, t_{i, a+1})\right)$
(with $t= t_{i, a} = t_{j, b}$).

In the following, we will refer to the above supra-adjacency representation as \textit{dyn-supra}.
We will moreover consider two variations of this representation.
First, we can ignore the direction of time of the original temporal network in the supra-adjacency representation
by making all links of $\mathcal{E}$ undirected.
We will refer to this representation as \textit{dyn-supra-undirected}.
Another possible variation consists in encoding the time delay between active nodes into edge weights,
with decreasing weights for increasing temporal differences.
This decay of edge weights is consistent with the idea that successive active nodes that are temporally far apart are less likely to influence one another (which is the case for many important dynamical processes).
In our case, we will consider, as a simple implementation of this concept, 
that the original weight of an edge 
$\left((i, t), (j, t')\right)$  
in the  \textit{dyn-supra} representation
is multiplied by the  reciprocal of the time difference between the active nodes, i.e.,
$\left|1/(t- t')\right|$. Each self-coupling edge
has thus weight 
$\omega/(t_{i, a+1}- t_{i, a})$, while a temporal
edge $(i,j,t)$ with $t=t_{i,a}=t_{j,b}$ yields the
edges 
$\left((i, t_{i, a}), (j, t_{j, b+1})\right)$ with
weight $w_{ij}(t) / (t_{j, b+1}- t_{i, a})$
 and 
 $\left((j, t_{j, b}), (i, t_{i, a+1})\right)$
 with weight 
 $w_{ij}(t) / (t_{i, a+1}- t_{j, b})$.
We will refer to this representation as \textit{dyn-supra-decay}.

\subsection{Node embedding}
The central idea of the node embedding method we propose for temporal networks,
which we call \textit{DyANE} (Dynamics-Aware Node Embeddings),
is to apply to the supra-adjacency network $\mathcal{G}$ any of the node embedding methods that have been developed for static networks.
Numerous embedding techniques have in fact been proposed 
for static networks, and we refer to the recent reviews~\cite{8294302,goyal2018graph}
for detailed descriptions. Most techniques consider as 
measure of proximity or similarity between nodes either a first-order proximity (two nodes are more similar if they are connected by an edge with larger weight) or a
second-order proximity (nodes are more similar
if their neighborhoods are similar). 
In particular, a popular way of exploring the (structural) 
similarity of nodes consists in using random walks
rooted at the nodes, which thus
explore their neighborhoods. Two of the most
well-known embedding techniques, 
DeepWalk \cite{Perozzi2014} and node2vec~\cite{Grover2016},
are based on such random walks. 

Methods based on random walks seem 
particularly appropriate in our framework.
Indeed, in the supra-adjacency representation, 
these random-walks will explore for each active node 
both the self-coupling edges leading to other
versions of the same original node, and the edges
representing the interactions between nodes.
As written above, these edges encode the paths along which dynamical processes occur, meaning that the final embedding will preserve structural similarities relevant to these dynamical processes. 
Here we will use DeepWalk~\cite{Perozzi2014},
as it is a simple and paradigmatic algorithm, and it is known to yield high performance in node classification tasks~\cite{goyal2018graph}.
Note that DeepWalk does not consider weighted edges,
but it can easily be generalized so that the random walks take into account edge weights~\cite{Grover2016}. 

\subsection{Prediction of dynamical states}
Once we have obtained an embedding for the supra-adjacency representation of the temporal network,
we can turn to the task of predicting the dynamical states of active nodes.
Since we assume that the set of possible states is known,
this is naturally cast as a (supervised) classification task,
in which each active node should be classified into one of the possible states.
In our specific case, the three possible node states are S, I, and R. We recall that the classification task is 
not informed by the actual dynamical process (except knowing the set of possible node states). In particular, no
information is available about the possible transitions 
nor about the parameters of the actual process.

We will use here a one-vs-rest logistic regression classifier, which is customarily 
used in multi-label node classification tasks based on embedding vectors. Naturally, we could use any other suitable classifier.
 
We remark  that we seek to predict active node states for individual realizations of the dynamics. This is relevant to several applications: for example, in the context of epidemic spreading, and given a temporal interaction network, 
one might use such a predictive capability to infer the state of all nodes from the observed states of few active nodes (``sentinel'' nodes).

\subsection{Evaluation}

The performance of our method can be evaluated along different lines. 
On the one hand, we can use standard measures used in prediction tasks, counting for each active
node whether its state has been correctly predicted.
We construct then a confusion matrix $C$, in which the element $C_{ss'}$ is given by the
number of active nodes that are in state $s$ in the  simulated spread and predicted to be in state $s'$ by the classification method. 
The number $TP_s$ of true positives for state $s$ is then the diagonal element $C_{ss}$ (and the total number of true positives
is $TP = \sum_s C_{ss}$), while the number of false negatives $FN_s$ for state $s$ is $\sum_{s' \ne s} C_{ss'}$.
Similarly, the number of false positives $FP_s$ is $\sum_{s'\ne s} C_{s's}$ (active nodes predicted to
be in state $s$ while they are in a different state in the actual simulation).

The standard performance metrics for each state $s$, namely precision and recall, are given respectively by $PRE_s = TP_s / (TP_s + FP_s)$ and
$REC_s = TP_s / (TP_s + FN_s)$ and the F1-score is $F1_s = 2 PRE_s \cdot REC_s / (PRE_s + REC_s)$.
In order to obtain overall performance metrics, it is customary to combine the per-class F1-scores 
into a single number, the classifier's overall F1-score. There are however several ways to do it and we 
resort here to the Macro-F1 and Micro-F1 indices, which  are widely used for evaluating 
multi-label node classification tasks~\cite{goyal2018graph}. 
Both indices range between 0 and 1, with higher values indicating better performance. 

Macro-F1 is an unweighted average of the F1 scores of each label, $\sum_{s \in {\mathcal S}} F1_s / |{ {\mathcal S}}|$.
On the other hand, Micro-F1 is obtained by using the total numbers of true and false positives and negatives. 
The total number of true positives is $TP = \sum_s C_{ss}$, and, since any classification error is both a false 
positive and a false negative, the total numbers of false positives and of false negatives are both equal to
$FP=FN=\sum_{s \ne s'} C_{ss'}$. As a result, Micro-F1 is $\sum_s C_{ss} / \sum_{s, s'} C_{ss'}$
(sum of the diagonal elements divided by sum of all the elements). In the case of imbalanced classes, 
Micro-F1 gives thus more importance to the largest classes,
while Macro-F1 gives the same importance to each class, whatever its size. In our specific case of the SIR model, 
the three classes S, I, R might indeed be very imbalanced, 
depending on the model parameters, so that it is important
to use both Macro- and Micro-F1 to evaluate the method's performance in a broad range of conditions.

From an epidemiological point of view, it is also interesting to focus on global measures corresponding
to an evaluation of the correctness of the prediction about the overall impact of the spread, as measured by the epidemic
curve and the final epidemic size. For instance, if we denote by $I_a^{real}(t)$ 
the numbers of infectious active nodes at time $t$ in the simulated spread, and by 
$I_a^{pred}(t)$ the number predicted in the classification task,
we can define as measure of discrepancy between the 
real and predicted epidemic curves:
\begin{equation}
\Delta_I = \frac{1}{T} \sum_{t=1}^T g(t), \nonumber \\
g(t) = 
\begin{cases}
0 &  \text{if $|V_t| = 0$} \\
\frac{| I_a^{pred}(t) - I_a^{real}(t)|}{|V_t|}  &  \text{otherwise}.
\end{cases}
\end{equation}
We can also focus on the final impact of the spread, as an evaluation of the global impact on the population,
and compute the discrepancy in the final epidemic size 
$$
\Delta_{size} = [(I^{pred}(T) + R^{pred}(T) )  - 
(I^{real}(T)+ R^{real}(T) )]/N \ .
$$
Note that not all nodes might be active at the last time
stamp $T$, so we can in this case and for simplicity 
consider for each node its last active time and assume that
it does not change state until $T$.

\subsection{Comparison with other methods and sensitivity analysis}

Our framework entails two
choices of procedures: the way in which the temporal network is represented as a static supra-adjacency object, and the choice of the node embedding method. 

First, we consider a variation of our proposed supra-adjacency representation (\textit{dyn-supra}),
using a ``baseline'' supra-adjacency representation, which we denote by \textit{mlayer-supra}:  in this representation,
we simply map each temporal edge $(i,j,t)$
to an edge between active nodes, namely 
$\left((i,t), (j, t)\right)$, similarly to the original supra-adjacency representation developed
for multilayer networks~\cite{Kivela:2014}.
Self-coupling edges are drawn as in \textit{dyn-supra}.

Moreover, for both \textit{dyn-supra} and \textit{mlayer-supra}, 
we considered an alternate embedding method to DeepWalk, 
namely LINE~\cite{Tang2015}, which embeds nodes in a way to preserve both first and second-order proximity.

In addition, we considered four state of the art embedding methods for temporal networks, which do not use the intermediate supra-adjacency representation, namely:
(i) DynamicTriad (DTriad)~\cite{zhou2018dynamic}, which embeds the temporal network by modeling triadic closure events; 
(ii) DynGEM~\cite{goyal2018dyngem}, which 
is based on a deep learning model. It outputs an embedding
for the network of each timestamp, initializing
the model at timestamp $t+1$ with the weights found at time $t$, thus transferring knowledge from $t$ to $t+1$ and learning about the changes from $G_t$ to $G_{t+1}$; 
(iii) StreamWalk~\cite{Beres2019}, which uses time-respecting walks and online machine learning to capture temporal changes in the network structure;
(iv) Online learning of second order node similarity
(Online-neighbor)~\cite{Beres2019}, which 
optimizes the embedding to match the 
neighborhood similarity of pairs of nodes,
as measured by the Jaccard index of these neighborhoods.

Overall, we obtain eight methods to create an embedding of the temporal network -- four variations of DyANE and four methods that directly embed temporal networks --
which we denote respectively \textit{dyn-supra}+DeepWalk, \textit{dyn-supra}+LINE, \textit{mlayer-supra}+DeepWalk, \textit{mlayer-supra}+LINE,
DTriad, DynGEM, StreamWalk and Online-neighbor.

Each variation of DyANE has moreover two parameters whose value can be a priori arbitrarily chosen, namely the weight $\omega$ and the embedding dimension $d$. In each of these variations of DyANE, it is also possible as explained above to consider undirected edges and to take into account the difference of the times between linked active nodes. 

For each obtained embedding, we will thus explore the performance of the classification task to explore the robustness of the results and their potential dependency on specific choices of the embedding method and of the parameter values.

\section{Numerical experiments and Results}
\label{sec:experiments}

In this section, we study the effectiveness of DyANE, in particular with the \textit{dyn-supra}+DeepWalk combination
to predict the nodes' epidemic states in a single instance of an
SIR spreading process.

To this aim, we use temporal networks built from empirical data sets that describe close-range proximity interactions of persons in a variety of real world environments.
We simulate the SIR (Susceptible-Infected-Recovered) dynamical process described above over these temporal networks, generating state labels for all active nodes.
Based on the above temporal networks and node labels, we run DyANE with different combinations of supra-adjacency representations and of embedding methods for the static network, and use the resulting embedding vectors as inputs to a supervised multi-label classifications tasks. 
We compare the results with the ones obtained with the 
other embedding methods described
in the previous section, and 
we test the sensitivity of our approach with respect to the choice of parameters and to the number of sampled active nodes ${D}$.

\subsection{Data sets and dynamical process}
We use publicly available 
data sets describing the face-to-face proximity of
individuals with a temporal resolution of 20 seconds~\cite{cattuto2010dynamics}. These data sets
were collected by the SocioPatterns collaboration\footnote{http://www.sociopatterns.org/}
and we specifically use data sets collected in 
offices ("InVS15"), a hospital ("LH10"), 
a highschool ("Thiers13"), 
a conference ("SFHH") and 
a school ("LyonSchool")~\cite{genois2018}. 
These data  correspond to a broad variety of contexts, 
with activity timelines, group structures and potential
correlations between structure and activity of different types.
 We built a weighted temporal network from each data set by
aggregating the data on $600$ seconds time windows.
Whenever multiple proximity events were registered
between two individuals within a time window, we
used the number of such events as the weight of the corresponding temporal edge.
Table~\ref{tab:num_temp_network} shows some basic statistics for each data set. 

\begin{table}[]
\centering
\caption{Empirical temporal networks.
Columns, from leftmost to rightmost: data set name, number of active nodes, number of nodes, number of timestamps, number of temporal edges, average weight of temporal edges, average fraction of timestamps in which a node is active.
\label{tab:num_temp_network} }
\scalebox{1}{
\begin{tabular}{|l|l|l|l|l|l|l|l}
\hline
 Name & $|\mathcal{V}|$ & $|V|$ & $|T|$ & $|E|$ & $\frac{1}{|E|} \sum_{e \in E} \, w(e)$ & $\frac{1}{|V| \, |T|}\sum_{v \in V} \, |\mathcal{V}_{v}|$\\ \hline
InVS15          & 22451    & 217                    & 699                & 37582              & 4.164           & 0.148                   \\
LH10            & 4880     & 76                     & 342                     & 14870              & 4.448      & 0.188                        \\
SFHH            & 10815    & 403                    & 127                    & 34446              & 4.079      & 0.211                        \\
Thiers13        & 32546    & 327                    & 246                    & 71724              & 5.256       & 0.405                       \\ 
LyonSchool      & 17174    & 242                    & 104                  & 89640              & 2.806         & 0.682                     \\ \hline
\end{tabular}
}
\end{table}

We simulated the SIR model on each such weighted temporal network, using the following five combinations 
of epidemic parameters: $(\beta,\mu)=$ $\{(0.25, 0.055)$, $(0.13, 0.1)$,
$(0.13, 0.055)$, $(0.13, 0.01)$, $(0.01, 0.055)\}$. 
In each case, we consider as initial 
state a single randomly selected node as seed, 
setting its state as infectious, with all others
susceptible. Given the stochastic nature of the model,
in some cases the infectious state barely spreads, 
with a large majority of the nodes remaining susceptible.
The prediction task would then be trivial, and we
restrict our study to non-trivial cases in which there
is still at least one infectious node when 
more than half of the total data set time span
has elapsed (i.e., $|I_{|T|/2}| \geq 1$). We thus run the SIR model up to $500$ 
times for each data set until we obtain a simulation in which the condition $|I_{|T|/2}| \geq 1$ is met. 
If this condition is not met in any of the $500$ simulations, we discard the corresponding case (see Table~\ref{tab:node_state_dist}).
For each selected simulation, we assign as ground truth label to each 
active node $(i,t_{i,a})$ the state of node $i$ at time $t_{i,a}$. 
Table~\ref{tab:node_state_dist} shows the proportions of each label among active nodes for each case.

\begin{table*}[]
\centering
\caption{
Proportions of the three possible states S, I and R among the active nodes in the runs considered, 
for each data set and each parameter set of the SIR model.
``-'' indicates the cases for which the epidemic did not spread sufficiently (i.e., $|I_{|T|/2}| = 0$).
\label{tab:node_state_dist}} 
\scalebox{1}{
\begin{tabular}{| l | l l l | l l l |  l l l |}
\hline
{} & \multicolumn{3}{l|}{$\beta=0.25$, $\mu=0.055$} & \multicolumn{3}{l|}{$\beta=0.13$, $\mu=0.1$} & \multicolumn{3}{l|}{$\beta=0.13$, $\mu=0.055$} \\
                 \cline{2-10} 
              Data set
                      & P(S)      & P(I)    & P(R)   & P(S)    & P(I)     & P(R)    & P(S)    & P(I)     & P(R)   \\
InVS15   & 0.064140     & 0.054652  & 0.88128  & -   & -  & -  & 0.104717   & 0.049129  & 0.846154   \\
LH10   &  0.136066   & 0.121311   & 0.742623 & 0.104098   & 0.084836   & 0.811066  & 0.077459   & 0.146721    & 0.775820  \\ 
SFHH   & 0.192510  &0.230236 & 0.577254  &  0.151826 &0.133703 &0.714471 & 0.271197 & 0.212575 & 0.516227 \\
Thiers13   & 0.193081 &0.087814 & 0.719105 & 0.172279 & 0.057242   & 0.770479   & 0.178271 & 0.094051 & 0.727678\\
LyonSchool  &0.183359 &0.188657& 0.627984 & 0.162397 & 0.102713 & 0.734890 & 0.196160 & 0.183999 & 0.619891  \\
\hline
\end{tabular}
}
\scalebox{1}{
\begin{tabular}{|l|lll|lll|lll|lll|lll|}
{}  & \multicolumn{3}{l|}{$\beta=0.13$, $\mu=0.01$} & \multicolumn{3}{l|}{$\beta=0.01$, $\mu=0.055$} \\
                 \cline{2-7} 
              Data set
                         & P(S)     & P(I)     & P(R)   & P(S)     & P(I)     & P(R)    \\ \hline
InVS15   & 0.192285    & 0.185782   & 0.621932    & 0.573917  & 0.041424  & 0.384660   \\
LH10     & 0.285246    & 0.339549  & 0.375205  & 0.323770 & 0.111680    & 0.564549   \\
SFHH   & 0.402312 &0.407582  &0.190106   & 0.810911 & 0.078132  & 0.110957    \\
Thiers13   & 0.468445 & 0.293062 & 0.238493 & 0.782892    & 0.046027  & 0.171081   \\ 
LyonSchool   & 0.161698 & 0.580703 & 0.257599 & 0.562536 & 0.154885 & 0.282578   \\ \hline
\end{tabular}
}
\end{table*}

We select uniformly at random $|D| =\rho |{V}|$ active nodes, and build our training set using those active nodes and the corresponding active node states.
Unless otherwise noted, $\rho=1$ (i.e., each node is observed on average once). 
We evaluate the prediction performance on a test data consisting of the remaining active nodes and their states.
We report the prediction performance averaged over five realizations of the embeddings and over five realizations of 
the random choice of training data, for each data set and parameter values.

\subsection{Implementation of the embedding methods}

We used publicly available implementations of all
embedding methods, namely the implementation of LINE\footnote{https://github.com/tangjianpku/LINE}, 
DynamicTriad\footnote{https://github.com/luckiezhou/DynamicTriad}, 
DynGEM\footnote{http://www-scf.usc.edu/$\sim$nkamra/}, 
StreamWalk and Online-neighbor
by the original authors \footnote{\label{StreamWalk}https://github.com/ferencberes/online-node2vec}. 
As for DeepWalk, we used an implementation of node2vec\footnote{https://github.com/aditya-grover/node2vec}
with $p=q=1$. Unless otherwise noted, we conducted experiments with
embedding dimension $d=128$ and self-coupling edge weight $\omega=1$. We used the 
default values of each implementation of the embedding methods,
except for the number of iterations of LINE, which we took equal to the number of samples of DeepWalk. 
We used Scikit-Learn~\cite{scikit-learn} to implement one-vs-rest logistic regression.

\subsection{Results}

\subsubsection{Prediction performance}

\begin{figure*}[h] 
    \centering
    \includegraphics[width=\linewidth]{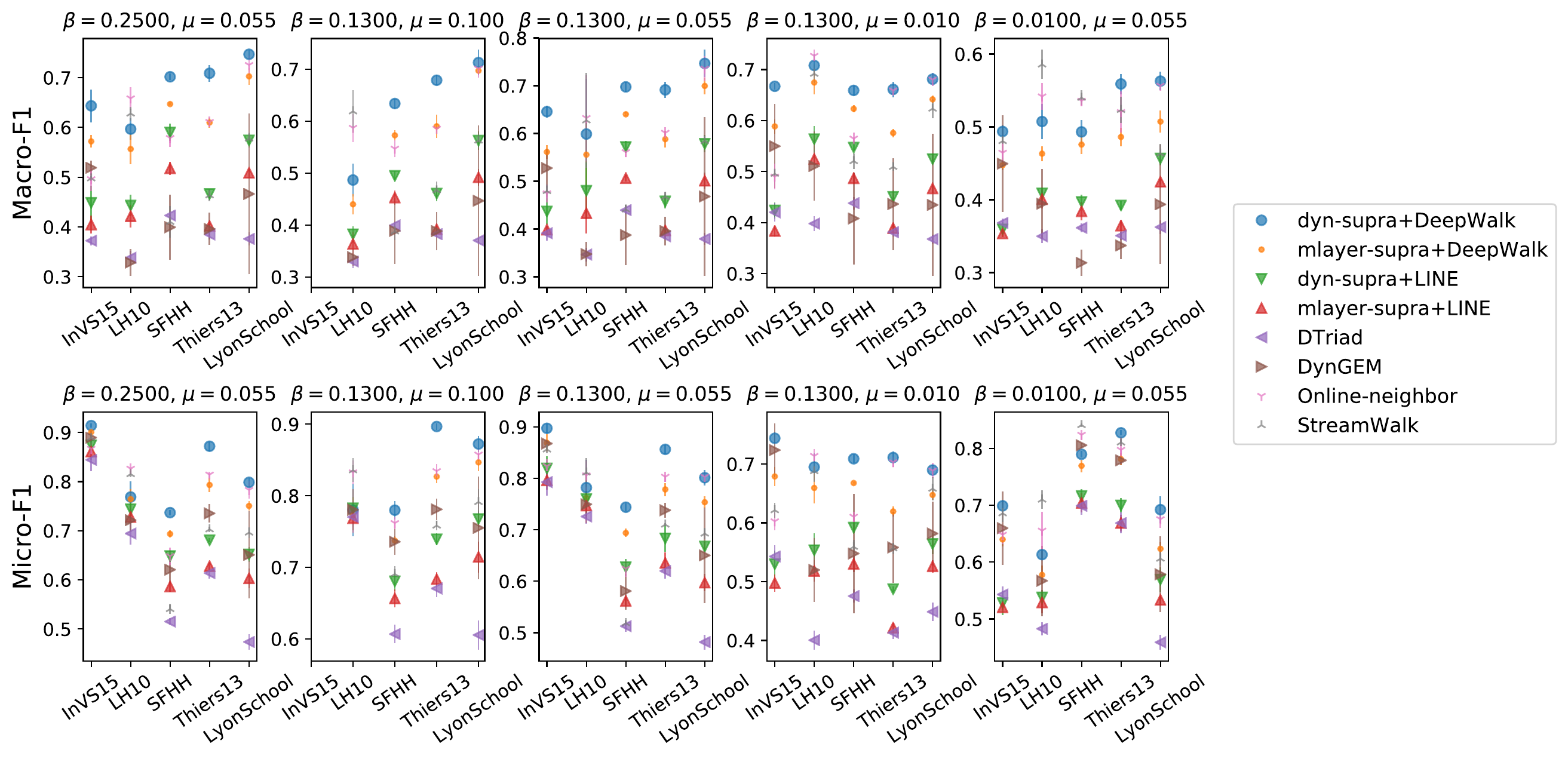} 
  \caption{Prediction performance of each method, for each data set and spreading parameter set. 
  The top and bottom row give the obtained values of Macro-F1 and Micro-F1, respectively.}
  \label{class_perf} 
\end{figure*}

Figure~\ref{class_perf} shows the prediction performance
of the eight methods considered, for all data sets
and SIR parameters considered. 
The \textit{dyn-supra} representation combined with DeepWalk yields almost always
the highest value both for Macro-F1 and Micro-F1, except for the LH10 data set (the smallest data set, see Table \ref{tab:num_temp_network}).
We moreover observe that:
(i) for a given static embedding method, the \textit{dyn-supra} supra-adjacency representation gives better results than the baseline (\textit{mlayer-supra}) one;
and (ii) for a given supra-adjacency representation, DeepWalk performs better than LINE.

\subsubsection{Epidemic impact and epidemic curves}

\begin{figure*}[h] 
    \centering
    \includegraphics[width=\linewidth]{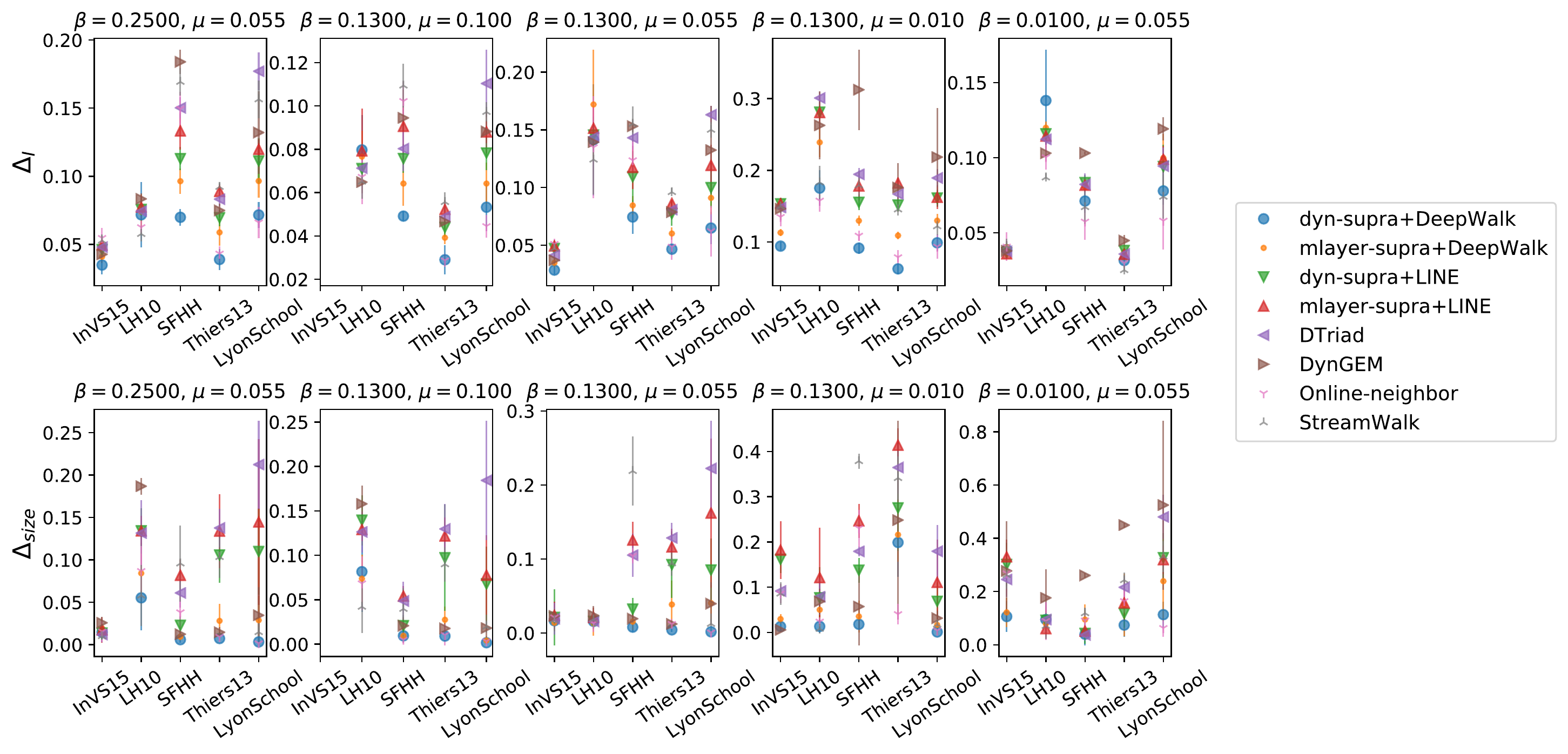} 
  \caption{Prediction performance of each method in terms of epidemic impact, for each data set and spreading parameter set. 
  The top and bottom row show respectively the cumulative discrepancy between the real and predicted epidemic curves $\Delta_I$ and
  the discrepancy between the final epidemic sizes $\Delta_{size}$.}
  \label{fig:deltasize_I} 
\end{figure*}

\begin{figure*}[h] 
    \centering
    \includegraphics[width=\linewidth]{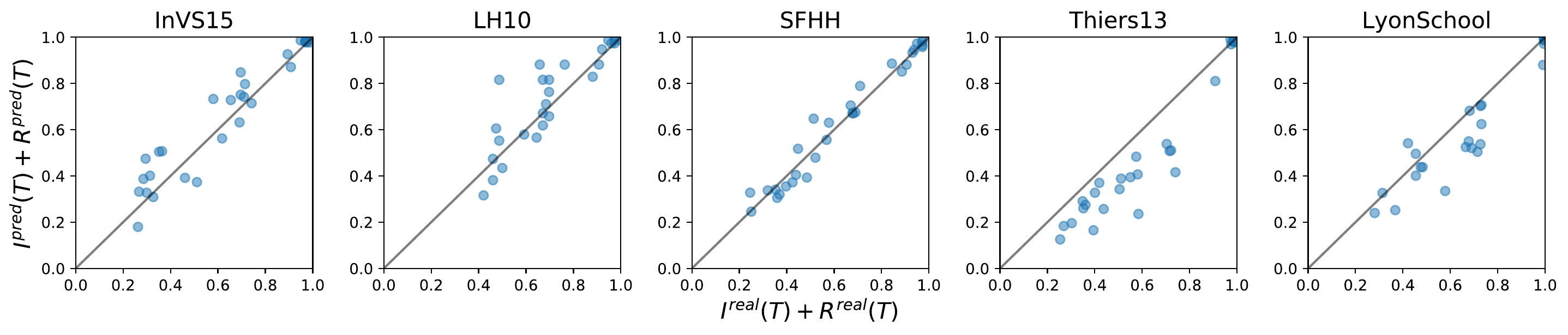} 
  \caption{Scatterplots of the final predicted epidemic size (number of infectious and recovered nodes at the end of the data set), vs. the real one. Each point corresponds
  to one simulation, and we vary the epidemic parameters in order to scan a wide range of outcomes. Here we use the \textit{dyn-supra}+DeepWalk method.
  The Pearson correlation coefficients for the predicted vs. real sizes are $0.957$ for InVS15, $0.88$ for LH10, $0.98$ for SFHH, $0.96$ for  Thiers13 and $0.95$ for LyonSchool.}
  \label{fig:scatter} 
\end{figure*}

Figure \ref{fig:deltasize_I} confirms the results of Fig. \ref{class_perf} from the point of view of the discrepancy measures between predicted and real epidemic curves
and from the point of view of the error in the predicted final epidemic size: $\Delta_I$ and $\Delta_{size}$ are small for most methods, data sets and parameter values, and
in particular the smallest error is most often obtained for the  \textit{dyn-supra}+DeepWalk method.
For this method, Figure \ref{fig:scatter} shows moreover scatterplots of the predicted vs. real final epidemic size, for all data sets: in each plot, epidemic parameters
have been varied to yield a large diversity of final epidemic sizes, showing that the predicted and real outcomes are very strongly correlated.

\begin{figure}[h] 
    \centering
    \includegraphics[width=\linewidth]{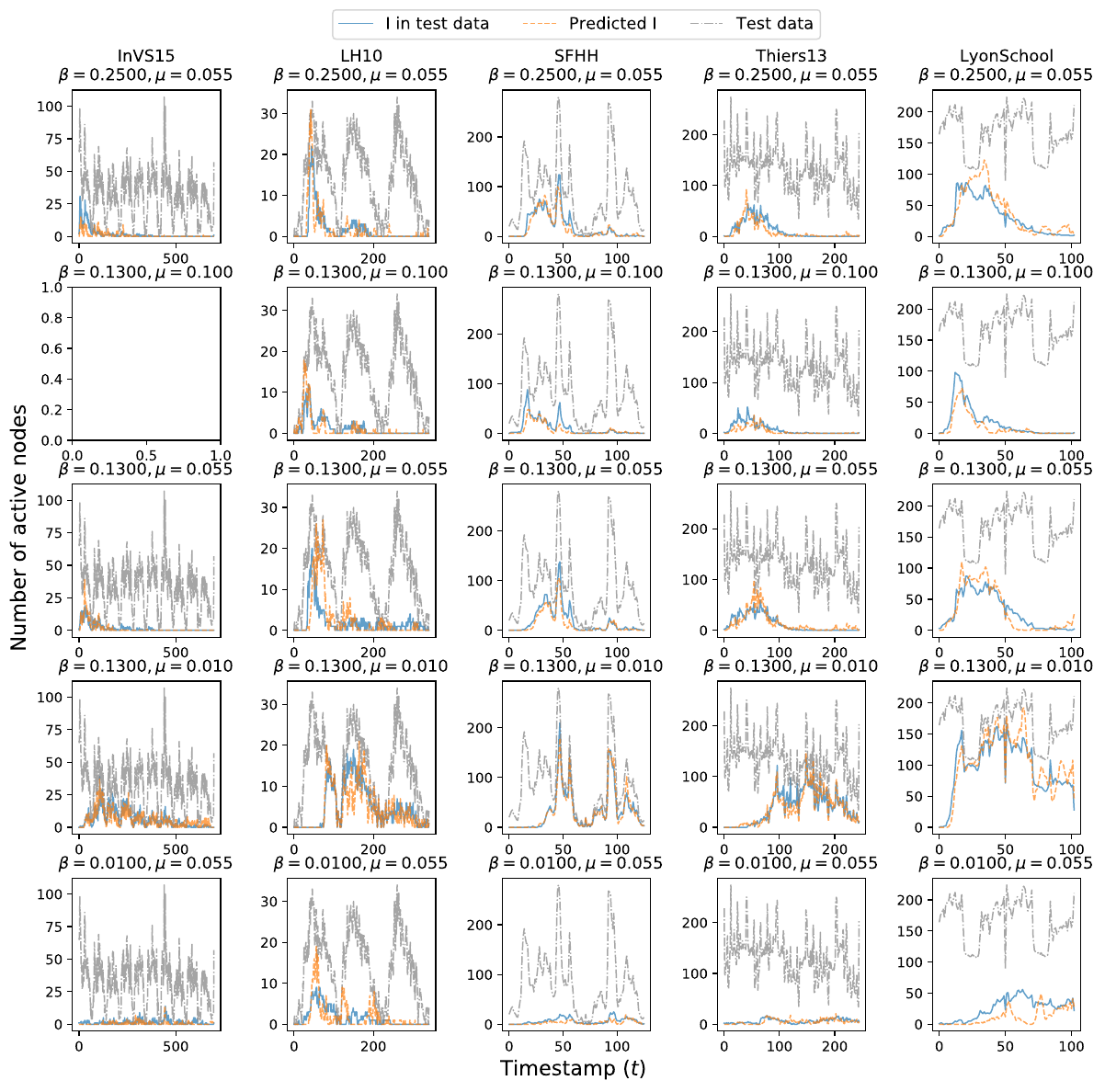} 
  \caption{Example of predicted timelines of the number of active nodes in the infectious state, for all data sets and $(\beta,\mu)$ parameter values,
  for the \textit{dyn-supra}+DeepWalk method.
    The blue, orange and gray lines are respectively
    the number of actual active nodes in state I in the test data, the number of predicted active nodes in state I and the number of active nodes in the test data at time $t$. 
    Note that the number of active nodes in the test data is almost the same as the total number of active nodes, as the training data is of small size ($\rho=1$, i.e., $|D|=|{V}|$).
    }
  \label{fig:epi_curves_alldatasets} 
\end{figure}

\begin{figure}[thb] 
    \centering
    \includegraphics[width=\linewidth]{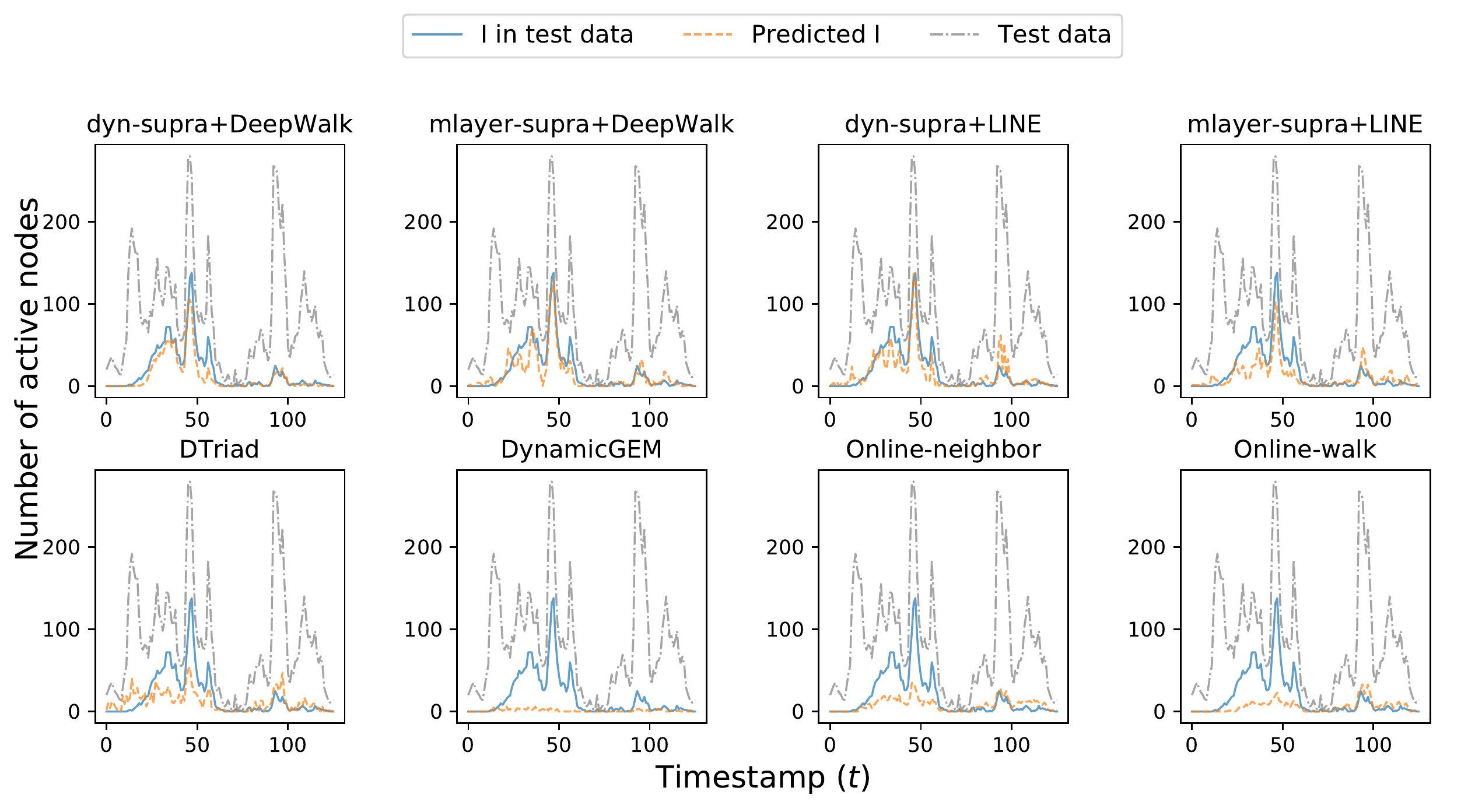} 
  \caption{Example of predicted timelines of the number of active nodes in the infectious state, for the SFHH (conference) data set
    for the various methods. Here $(\beta,\mu)=(0.13,0.055)$ and $\rho=1$. Line colors have the same meaning as in Fig. \ref{fig:epi_curves_alldatasets}.
  }
  \label{fig:epi_curves_SFHHallmethods} 
\end{figure}

Figures \ref{fig:epi_curves_alldatasets} and \ref{fig:epi_curves_SFHHallmethods} give a more qualitative illustration of the performance of 
our method in the reconstruction of the epidemic curves, highlighting as well the capacity of the method to recover the timing of epidemic peaks.
This is particularly relevant, as heights and timings of peaks in the number of infectious determine the eventual burden on the healthcare system.
Figure \ref{fig:epi_curves_alldatasets} first shows that the  \textit{dyn-supra}+DeepWalk method recovers well the periods of large and small number of infectious
individuals for all data sets and over a wide range of parameter values. Moreover, Figure \ref{fig:epi_curves_SFHHallmethods} shows 
that the four methods combining a supra-adjacency representation
with either DeepWalk or LINE yield good results, while the four other methods strongly underestimate the largest epidemic peak, predicting epidemic curves that 
spread out  the epidemic impact more evenly over the whole timeline, and thus yielding a less accurate information in terms of size and timing of the 
largest epidemic burden. 

\clearpage
\newpage

\subsubsection{Sensitivity analysis}

\begin{figure}[t] 
    \centering
    \includegraphics[width=0.7\linewidth]{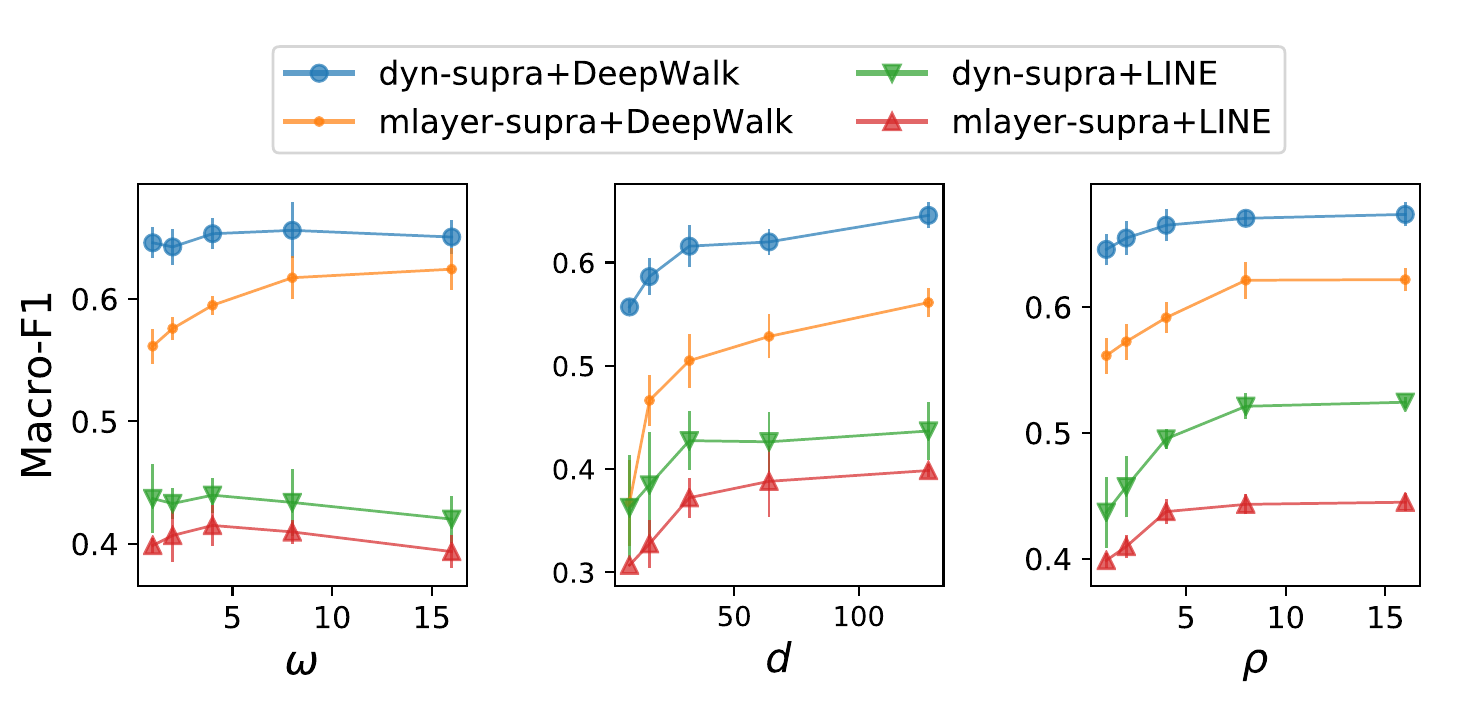}       
        \includegraphics[width=0.7\linewidth]{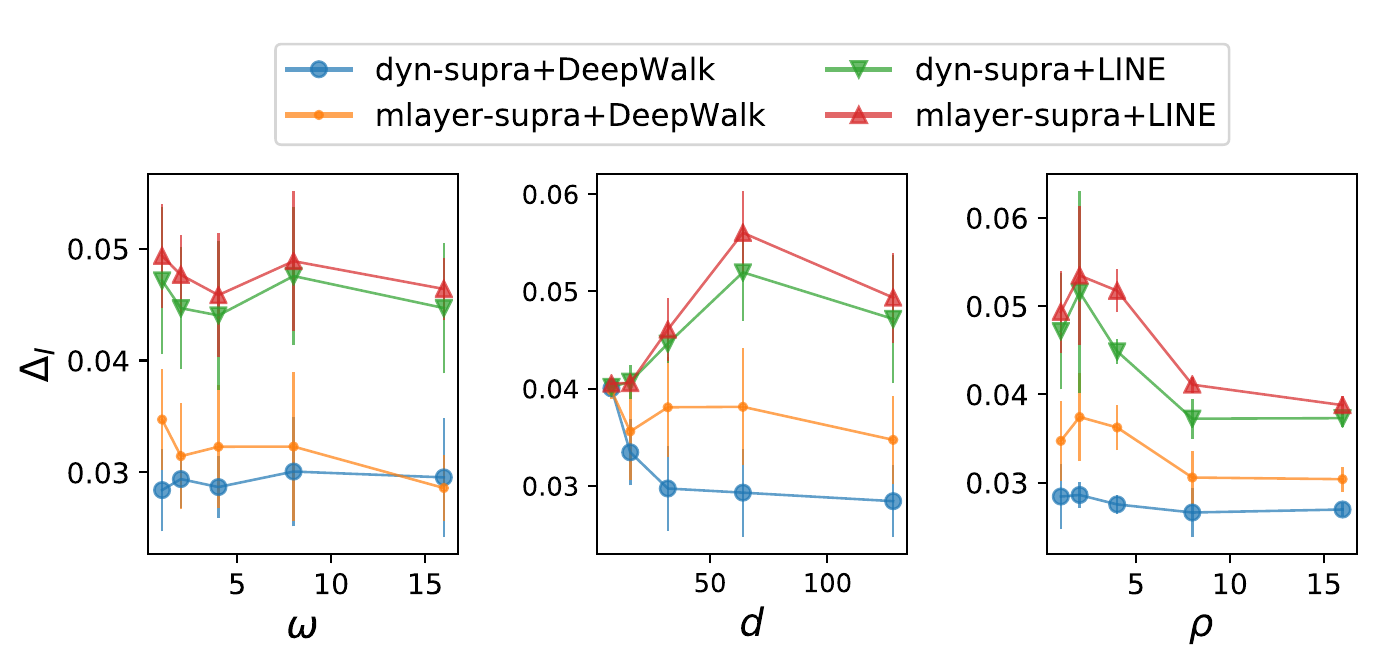}     
  \caption{Results of the hyper-parameter sensitivity on the Macro-F1 and $\Delta_I$ performace measures for the InVS15 data set. Bars indicate the
  standard deviation. Here $(\beta, \mu) = (0.13,0.055)$.} 
    \label{param_sensitivity:a}     
\end{figure}

We now investigate the effect of the hyper-parameters of the supra-adjacency representation 
(the weight $\omega$ of self-coupling edges) and of the embedding (the embedding dimension $d$). 
We  show in Fig.~\ref{param_sensitivity:a} 
the results obtained for two performance measures, for the InVS15 data set and
$(\beta, \mu) = (0.13,0.055)$, but we have confirmed
the same tendency for the other data sets, parameter values and for the Micro-F1 and $\Delta_{size}$ measures. The results show
that the performance of \textit{dyn-supra}+DeepWalk is very stable
with respect to changes in $\omega$. The performance 
is also stable on a wide range of embedding dimensions, and
decreases when it becomes smaller than $\approx 50$. Overall,
\textit{dyn-supra}+DeepWalk remains very effective without the need for fine-tuning $\omega$ or $d$.

Figure~\ref{param_sensitivity:a} also shows the effect
of increasing the parameter $\rho$, i.e., of
being able to observe a larger fraction of active nodes.
The performance slightly increases with $\rho$ and in particular \textit{dyn-supra}+DeepWalk consistently yields the best result at all values of $\rho$.

\begin{figure}[t] 
   \centering
    \includegraphics[width=0.7\linewidth]{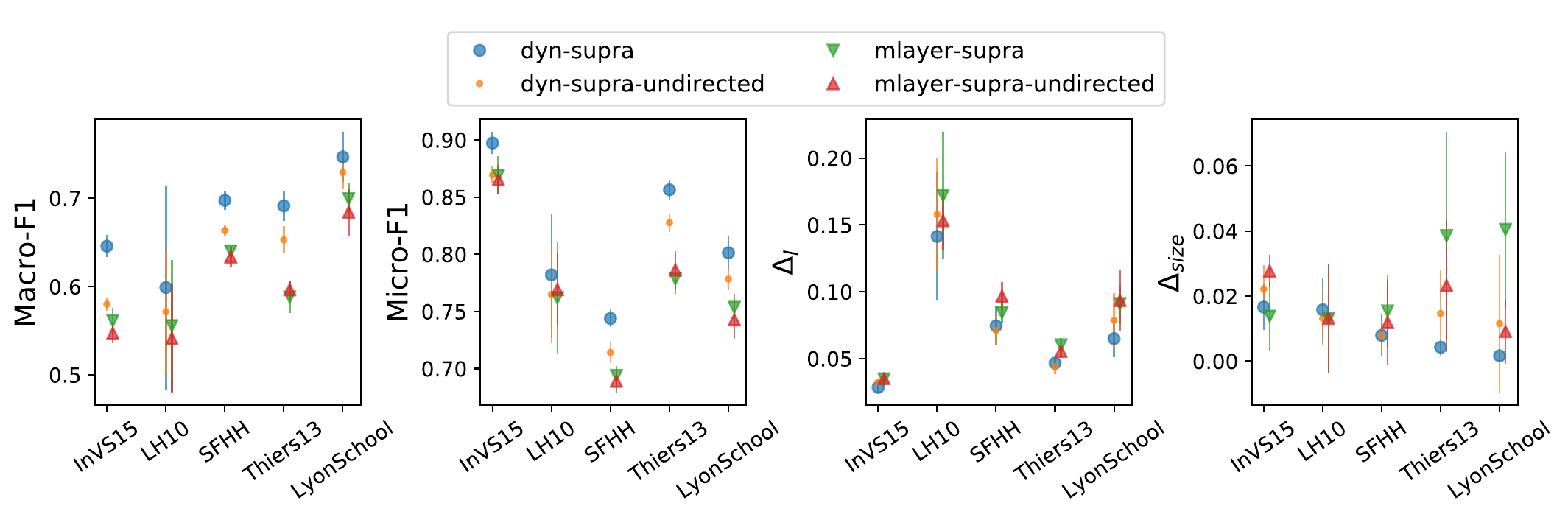} 
    \label{temp_dist} 
    \includegraphics[width=0.7\linewidth]{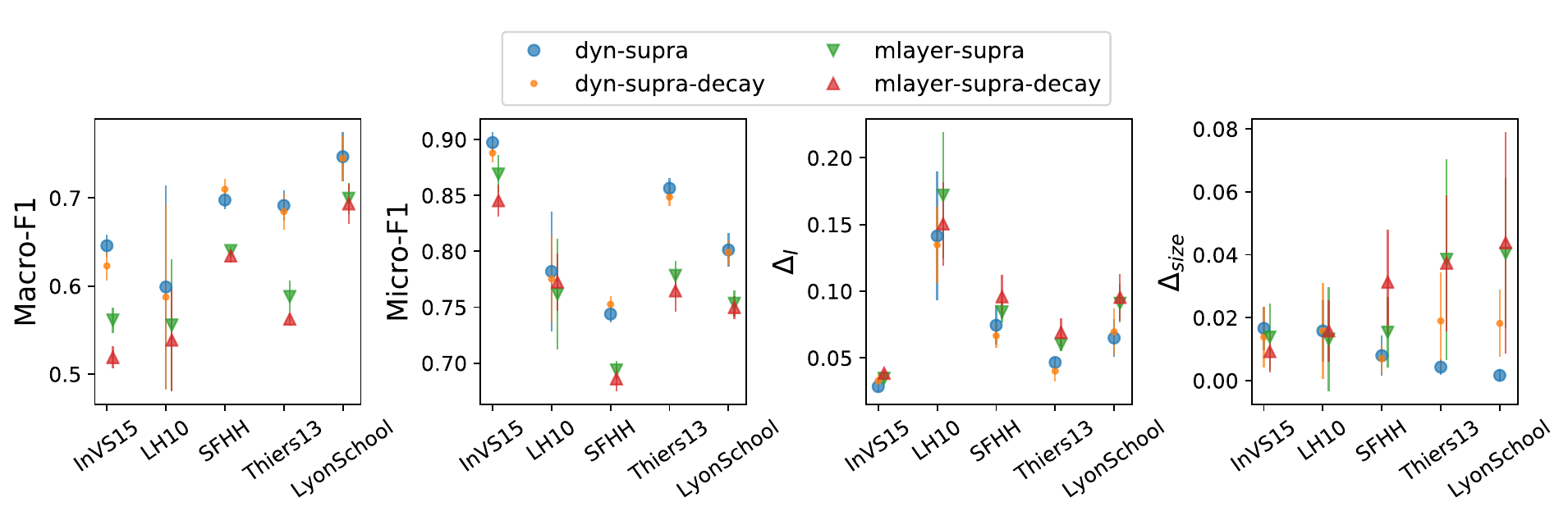} 
    \label{temp_dir} 
  \caption{Effect of the variations in the supra-adjacency representation, for the DeepWalk embedding. Here $(\beta, \mu) = (0.13, 0.055)$.
  Top: effect of using undirected edges.
  Bottom: effect of introducing weights depending
  on time difference.}
  \label{temp_inter_edge} 
\end{figure}

As mentioned above, we finally consider two variations if the 
\textit{dyn-supra} representation:
(i) we regard edges as directed towards increasing timestamps (\textit{dyn-supra-directed});
(ii) we let the weight of an edge decay with increasing temporal lag between the active nodes it links, e.g., we modulate the edge weight according to the reciprocal of the lag (\textit{dyn-supra-decay}).
We also consider these variations for \textit{mlayer-supra} representation, yielding \textit{mlayer-supra-directed} and \textit{mlayer-supra-decay}, respectively. 
Notice that, in the \textit{mlayer-supra} method, the supra-adjacency edges representing temporal edges are actually not affected by these variations.
We report in Fig.~\ref{temp_inter_edge}  the results for $(\beta, \mu) = (0.13, 0.055)$
and for the DeepWalk embedding, as DeepWalk overall yielded the best results.
We checked that the results of Fig.~\ref{temp_inter_edge} hold similarly for the LINE  embeddings.

Figure~\ref{temp_inter_edge} indicates that using undirected
edges slightly worsens the performance of both \textit{dyn-supra} and \textit{mlayer-supra}
methods. Introducing weights that depend on the time
difference between active edges also worsens
the performance for \textit{mlayer-supra}, with little effect
on \textit{dyn-supra}. Overall, the original \textit{dyn-supra} method with
directed edges and using only the weights of the original
temporal edges yields the highest prediction performance.

\section{Conclusion}
\label{sec:conclusion}

We have introduced a new method to recover  the 
dynamical evolution of a single instance of a process occurring on a temporal network, 
from partial observations and without information on the nature of the process itself except
from the set of possible states of the nodes. 
Our strategy is based on leveraging the field of node embedding techniques and on the introduction of a new
method for embedding nodes of temporal networks aimed at providing low-dimensional feature 
vectors that are informative of dynamical processes occurring over temporal networks.

Our method first maps the temporal
network to a modified supra-adjacency representation, 
which preserves the paths on which the process
unfolds. As this representation yields a static
graph among the active nodes, which are pairs 
of the form (node of the temporal network, time of interaction),
it enables  the use of embedding techniques for static networks.
We choose to use DeepWalk, as it is a simple and paradigmatic algorithm based on random walks and thus  
particularly suited to explore the neighborhood
of the nodes of the supra-adjacency representation in
a way relevant to the dynamical
process on the network. We finally frame the 
inference of the dynamical state of all active nodes
from a set of observations as a supervised classification task.

We have shown the performance of our method 
on the case of an epidemic-like model
on empirical temporal networks and compared it
with other state of the art methods. Our method
consistently yields very good classification performance
in a robust way across data sets and process parameters,
without fine-tuning hyper-parameters.

Our results show that it is possible, without any knowledge of the precise nature of the process nor of
its parameters, to recover crucial information on
its outcome, even with a very limited number of observations (for most of our results, each node is observed on average once). 
Note in particular that our method assumes no knowledge of which transitions between states actually
occur in the real dynamics: this means that the predicted sequence of states of each individual node might yield
"forbidden" transitions (e.g., in the SIR example, transitions from I to S or from R to I). Nevertheless, we have shown that
the outcome of the classification task gives a good estimation of the actual dynamics, as quantified both by usual measures of prediction task
performance and measures focusing more on the epidemic burden, such as the cumulative discrepancy between predicted and real epidemic curves
and the difference between predicted and real final epidemic sizes. We have also shown that the height and timing of the epidemic curve, which {\em in fine}
determines the period of worst expected burden on the healthcare system during an epidemic, are also well reproduced in our framework, while the other 
embedding methods for temporal networks predict a more spread out epidemic over the whole temporal window, with an underestimation of the epidemic peak height.

Our method has the clear limitation that we assume 
the whole temporal network to be known.
Although a full observation of the contact patterns 
of individuals could be envisioned in some specific controlled settings such as hospitals, this is not
generally the case. Further work will 
address this limitation by considering the effect of 
noise and errors in the temporal network data, and by considering the case in which only a (more or less detailed) set of statistics of the temporal network is known. 
Noise could also impact the quality of
the sampling (e.g., observational errors), and
we will check its impact on our method's performance.
Further work will also address different sampling strategies
such as a sampling concentrated at early times, or focused on few
specific ``sentinel'' nodes followed at all times,
or of a whole snapshot of the system but only at a specific time. 
This could yield interesting insights on how to optimize surveillance strategies in concrete settings.

Finally, since our method is largely agnostic with respect to the specific dynamical process, we will consider other processes such as other models of disease propagation, complex contagion
phenomena or opinion formation.

\section*{Acknowledgments}
This study was partially supported by the Lagrange Project of the ISI Foundation funded by CRT Foundation to CC.
It was partially supported by the ANR project DATAREDUX (ANR-19-CE46-0008-01) to AB.

\bibliography{references}

\end{document}